\documentclass[preprint,aps]{revtex4}
\usepackage{color}
\usepackage{amsmath,amssymb,amsfonts,dcolumn,color,graphicx,graphics,latexsym,epsfig}
\def\beq{\begin{equation}}
\def\eeq{\end{equation}}
\def\barr{\begin{array}}
\def\earr{\end{array}}

\begin{document}

\title{A note on spherically symmetric, static spacetimes in Kanno-Soda 
on-brane gravity}

\author{Sayan Kar${}^\#$, Sayantani Lahiri${}^{\dagger,\ddagger}$ and 
Soumitra SenGupta ${}^{*}$}
\email{sayan@iitkgp.ac.in,sayantani.lahiri@gmail.com,tpssg@iacs.res.in}
\affiliation{${}^\#$Department of Physics {\it and} Center for Theoretical Studies \\Indian Institute of Technology, Kharagpur, 721 302, India,}
\affiliation{${}^{\dagger}$Institute for Physics, University Oldenburg, D-26111 Oldenburg, Germany,}
\affiliation{${}^{\ddagger}$ZARM, University of Bremen, Am Fallturm, 28359 Bremen, Germany}
\affiliation{${}^{**}$Department of Theoretical Physics, Indian Association for the
Cultivation of Science
\\ 2A and 2B Raja S.C. Mallick Road, Jadavpur, Kolkata 700 032, India.}

\begin{abstract}
\noindent  
Spherically symmetric, static on-brane geometries 
in the Kanno-Soda (KS) effective scalar-tensor theory of on-brane 
gravity are discussed. In order to avoid brane collisions and/or
an infinite inter-brane distance, at finite values of the brane coordinates,
it is necessary that the radion scalar be everywhere finite and non-zero.
This requirement constrains the viability of the standard, well-known 
solutions in General Relativity (GR), in the context of the KS effective 
theory. The radion for the  Schwarzschild solution does not satisfy
the above requirement. 
For the Reissner--Nordstrom (RN) naked singularity and the extremal
RN solution, one can obtain everywhere finite, non-zero
radion profiles, though the required on-brane matter violates the 
Weak Energy Condition. In contrast, for the RN black hole, 
the radion profile yields a divergent inter-brane distance at the horizon, 
which makes the solution unphysical. 
Thus, both the Schwarzschild  
and the RN solutions can be meaningful in the KS effective theory, 
only in the trivial GR limit, i.e. with a constant, non-zero radion. 
\end{abstract}

\pacs{}

\maketitle

\newpage

\section{Introduction}

\noindent Effective, on-brane theories of gravity have been
in vogue ever since the Randall-Sundrum warped braneworld
scenario was proposed\cite{randall}. Among such four dimensional
gravity theories, the most well-known one is due to
Shiromizu, Maeda and Sasaki (SMS) \cite{sms} which considers a bulk with an
infinite extra-dimension and a single brane. There have been
proposals on effective theories in a two-brane set-up. 
In this article, we consider one such effective theory due to Kanno and 
Soda (KS) 
\cite{kanno}. A major difference between the SMS and the KS effective theories
is the presence of a non-local (bulk dependent) term in the former
and the absence of any non-locality in the latter.
Our objective is to look for solutions in the KS
effective theory, keeping in mind that the radion field, which is
linked to the distance between the branes, (i) is 
never zero in value (thus, avoiding brane collisions) and (ii) 
does not diverge at any finite value of the brane coordinates.

\noindent Recently \cite{kls}, we have discussed some cosmological and 
spherically 
symmetric, static spacetimes in this four dimensional, effective, 
on-brane, scalar-tensor theory of gravity. The spherically symmetric, static
solutions obtained in  
\cite{kls} turned out to be the Majumdar-Papapetrou
solution \cite{mp} with the source being the effective 
scalar field (radion) energy-momentum and additional on-brane matter. 
Here, we ask a broader question: are the standard General Relativity (GR)
solutions like the Schwarzschild or the
Reissner--Nordstrom (RN) permissible in the KS theory? Obviously, we
do not expect the GR solutions to arise in the KS theory with the same
matter content as in GR. Rather, we would like to find out if 
the radion energy momentum and
extra on-brane matter can conspire in unison to allow the Schwarzschild 
or the RN solution in the KS effective theory.    

\noindent It should be noted that there are several aspects related to this
question, a couple of which we have already mentioned. 
In addition to having a radion which is finite and non-zero everywhere,
the  on-brane matter  must also be physically reasonable
in the classical sense, i.e. it must satisfy one of the well-known energy
conditions, such as the Weak Energy Condition or the Null Energy Condition
\cite{wald}.

\noindent We may recall that the
Reissner-Nordstrom solution does arise as a solution \cite{rn} of the 
Shiromizu-Maeda-Sasaki (SMS) single brane effective Einstein equations \cite{sms}, where
the non-local contribution from the bulk Weyl tensor (the traceless 
${\cal E}_{\mu\nu}$)
acts as its source without any explicit on-brane matter. 
The functional form of the ${\cal E}_{\mu\nu}$
depends on the bulk Weyl tensor and other features of
the bulk geometry.  It cannot be determined uniquely from the knowledge 
of four dimensional, on-brane physics.  
In contrast, in the KS effective theory, the influence of the bulk is
exclusively through the radion field which depends only on the brane 
coordinates.
It is therefore meaningful to ask whether the equations which
arise in the effective, on-brane Kanno-Soda theory (which are local and 
different from those obtained in the single brane SMS effective theory) 
also admit a RN or a Schwarzschild solution, in some way.  

\noindent We will mainly work with the Reissner--Nordstrom
solution written in isotropic coordinates.  After obtaining the
radion profiles in the various cases, we will 
see if the radion satisfies the necessary requirements. 
Subsequently, we will analyse the nature of required on-brane 
matter with reference to the Weak Energy Condition \cite{wald}.

\section{The Kanno-Soda effective theory: main equations}

\noindent 
The effective on-brane scalar-tensor theories developed by Kanno and Soda 
\cite{kanno} in the context of the Randall--Sundrum two-brane model 
leads to the following Einstein-like equations on the 
visible `b' brane \cite{kanno},
\begin{eqnarray}
G_{\mu \nu} = \frac{\kappa^2}{l\Phi} T^{b}_{\mu \nu} + \frac{\kappa^2\,(1 + \Phi)}{l\Phi} 
T^{a}_{\mu \nu} + \frac{1}{\Phi}\left ({\nabla}_{\mu} {\nabla}_{\nu} \Phi - g_{\mu \nu} {\nabla}^{\alpha} {\nabla}_{\alpha} \Phi\right ) \nonumber 
\\  -\frac{3}{2\Phi(1+\Phi)} \left ({\nabla}_{\mu}\Phi{\nabla}_{\nu}\Phi
-\frac{1}{2} g_{\mu \nu} {\nabla}^{\alpha}\Phi{\nabla}_{\alpha} \Phi\right )  \label{EE}
\end{eqnarray}
Here $g_{\mu\nu}$ is the on-brane metric, the covariant
differentiation is defined with respect to  $g_{\mu\nu}$ and we have taken the 
five dimensional line  element as,
\begin{eqnarray}
ds_5^2= e^{2\phi(x)} dy^2 + {\tilde g}_{\mu\nu} (y, x^{\mu}) dx^\mu dx^\nu
\end{eqnarray}
$\kappa ^2$ is the $5D$ gravitational coupling constant. 
$T^{a}_{\mu \nu}$, $T^{b}_{\mu \nu}$ are the
stress-energy on the Planck brane and the visible brane respectively. 
The appearance of $T^{a}_{\mu\nu}$ (matter energy momentum on the `a' brane)
in the field equations on the `b' brane, inspired the usage of the term
`quasi-scalar-tensor theory'. However, if we assume $T^a_{\mu\nu}=0$ then we
have the usual scalar-tensor theory.

\noindent We denote $d(x)$ as the proper distance 
between branes located at $y=0$ and $y=l$. $d(x)$ is defined as,
\beq
d(x)\,=\, \int^{l}_{0}e^{\phi (x)} dy
\eeq
We further define $\Phi = e^{2\frac{d}{l}}-1 $.
It may be observed that the viability of such a model with a everywhere
finite and non-zero brane
separation, implies that (i) the minimum of $d(x)$ is not equal to zero 
and (ii) $d(x)$ is never infinity at any finite value of the
brane coordinates. This, in turn indicates that the value of $\Phi(x)$ is
always greater than zero and $\Phi(x)$ never becomes infinity. 

\noindent Note that the radion contribution
on the R. H. S. of the field equation is traceless, which is reminiscent 
of the traceless ${\cal E}_{\mu\nu}$ in the SMS effective theory.
 
\noindent The scalar field equation of motion on the visible brane is 
given as,
\begin{equation}
{\nabla}^{\alpha}{\nabla}_{\alpha}\Phi =\frac{\kappa^2}{l}\frac{T^a + T^b}{2\omega+3} - 
\frac{1}{2\omega +3} \frac{d\omega}{d\Phi} ({\nabla}^{\alpha}\Phi)(
{\nabla}_{\alpha}\Phi )
\end{equation}
where $T^{a}$, $T^{b}$ are the traces of energy momentum tensors on 
Planck (`a') and visible (`b') branes, respectively. 
The coupling function $\omega({\Phi})$ expressed in terms of $\Phi$ is,
\begin{equation}
\omega (\Phi) = -\frac{3\Phi}{2(1+\Phi)}
\end{equation}
We know that gravity on both the branes are not independent. 
Dynamics on the Planck brane at $y=0$ is linked to that on the visible brane 
through the relation \cite{kanno} :
\beq
\Phi(x)\,=\,\frac{\Psi}{1 - \Psi }
\eeq
where $\Psi$ is the radion field as defined on Planck brane. 
The induced metric on the visible brane can be expressed in terms of 
$\Psi$ as,
\beq
g^{b-brane}_{\mu \nu}\,=\,(1 - \Psi)\,[h_{\mu \nu} + g^{(1)}_{\mu \nu}(h_{\mu \nu},\Psi, T^{a}_{\mu \nu},T^{b}_{\mu \nu},y=l)]
\eeq
where $g^{(1)}_{\mu \nu}$ is the first order correction term (see 
\cite{kanno} for details). It is possible to work with the gravity theory
and the $\Psi$ field equation on the `a' brane.  

\noindent In our work here, we assume that the on-brane stress
energy is nonzero only on the `b' brane (visible brane). 
We also assume that the on-brane matter is traceless and therefore,
since the effective radion stress energy is also traceless,
the Ricci scalar of the spacetime geometry is identically zero. 
Conversely, if we assume $R=0$, the on-brane matter is traceless. This choice of
$R=0$ enables us to propose the standard General Relativity solutions
(like Schwarzschild and Reissner-Nordstrom) as solutions in the
Kanno-Soda effective theory with on-brane matter. The two main
hurdles we need to address are therefore: 

$\bullet$ Are the standard GR solutions also solutions in the
effective theory, with a non-zero, everywhere finite radion?

$\bullet$ What is the nature of the on-brane matter 
required to support such standard GR solutions?

\section{Spherically symmetric, static solutions}

\subsection{Line element, field equations and the radion}

\noindent 
Let us assume a four dimensional line element on the visible brane, in
isotropic coordinates, given as 
\begin{equation}
ds^2=-\frac{f^2(r)}{U^2(r)} dt^2 +U^2(r) \left [ dr^2 + r^2 d\theta^2
+r^2\sin^2\theta d\phi^2 \right ]
\end{equation}
where $U(r)$ and $f(r)$ are the unknown functions to be determined from the
Einstein-like equations.
Using the above line element ansatz and the assumption that 
$\Phi$ is a function of $r$ alone, we get the following field equations
from the Einstein field equations mentioned above.
\begin{eqnarray}
-2 \frac{U''}{U} +\left (\frac{U'}{U}\right )^2 - 4\frac{U'}{Ur} =
-\frac{\Phi'^2}{4\Phi(1+\Phi)} + \left (\frac{U'}{U} -\frac{f'}{f}\right ) 
\frac{\Phi'}{\Phi} +\frac{\kappa^2}{l \Phi} \rho  \\
-\left (\frac{U'}{U}\right )^2 + 2\frac{f'}{f} \left (\frac{U'}{U} + \frac{1}{r}\right ) =  
-\frac{3\Phi'^2}{4\Phi(1+\Phi)} - \frac{U'}{U}\frac{\Phi'}{\Phi} -
\frac{2\Phi'}{\Phi r} - \frac{f'}{f} \frac{\Phi'}{\Phi} +\frac{\kappa^2}{l\Phi} \tau \\
\left (\frac{U'}{U}\right )^2 + \frac{f''}{f} -2\frac{f'}{f} \frac{U'}{U} +\frac{f'}{f}\frac{1}{r}  =  
\frac{\Phi'^2}{4\Phi(1+\Phi)} + \frac{U'}{U}\frac{\Phi'}{\Phi} +
\frac{\Phi'}{\Phi r} +\frac{\kappa^2}{l\Phi} p
\end{eqnarray}
where $\rho$, $\tau$ and $p$ correspond to on-brane matter and
using the tracelessness condition we have $-\rho+\tau +2 p =0$.
We have absorbed a factor of $U^2$ in the definitions of $\rho$,
$\tau$ and $p$.

On the other hand, the scalar ($\Phi$) field equation gives
\begin{equation}
\Phi'' + \frac{f'}{f} \Phi' + 2\frac{\Phi'}{r} = \frac{\Phi'^2}{2(1+\Phi)}
\end{equation}
which can be integrated once to get
\begin{equation}
\frac{\Phi'}{\sqrt{1+\Phi}} = \frac{2C_1}{r^2 f}
\end{equation}
where $C_1$ is a positive, non-zero constant. It is useful to note
here that the radion depends only on the metric function $f(r)$ and not on 
$U(r)$.
We also know that the existence of a horizon in a static, spherically
symmetric geometry is linked with the
existence of zeros in $f(r)$. Thus, $\Phi'$ will always diverge at 
the horizon of any spherically symmetric, static spacetime.

\noindent The field equations with 
the requirement of traceless on-brane matter, leads to the 
following equation for the metric function $f$ and $U$:
\begin{equation}
\frac{U''}{U} + \frac{f''}{f} -\frac{f'}{f}\frac{U'}{U} + 2\frac{f'}{f r}
+ 2 \frac{U'}{U r} =0
\end{equation}
\noindent A solution for $f$ and $U$ which satisfies the above tracelessness
condition can be found by recalling the 
Reissner--Nordstrom solution written in isotropic coordinates.
For such a solution we have,
\begin{eqnarray}
f(r)=1-\frac{M^2}{4 r^2} + \frac{e^2}{4 r^2} \\
U(r) = 1+\frac{M}{r} +\frac{M^2}{4 r^2} -\frac{e^2}{4r^2}
\end{eqnarray}
where $M^2$ and $e^2$ are constants. We have retained the notation
of `$e$' and `$M$' used in the standard GR Reissner-Nordstrom solution
where they represent charge and mass, respectively. However,
here, `$e$' and `$M$' may not carry the same physical meaning as
in Reissner-Nordstrom.  
It is easy to check that the above-written functional forms of $f$ and
$U$ satisfy the tracelessness criterion.

\noindent We now look at the equation for the scalar $\Phi$. Assume
$1+\Phi = \xi^2$. The first integral of the scalar wave equation
then becomes
\begin{equation}
\xi' = \frac{C_1}{r^2 f} = \frac{C_1}{r^2 + a^2}
\end{equation}
where $a^2 = \frac{e^2-M^2}{4}$ and $C_1$, an integration constant. 
It is clear that there will be
two different solutions for $a^2>0$ ($e^2>M^2$) and $a^2<0$ ($e^2< M^2$).
Both these solutions must converge to the solution for $e^2=M^2$
which gives the extremal limit. 

\noindent When $e^2>M^2$ (i.e. the nakedly singular solution) we obtain
\begin{equation}
\Phi (r) = \left ( \frac{C_1}{a}\tan^{-1} \frac{r}{a} + \frac{C_4}{2}\right )^2 -1
\end{equation}
This solution remains valid for all $r\ge 0$ with a choice 
of $C_4 > 2$ necessary to ensure the positivity condition $\Phi(r)>0$. 
In addition, note that the radion is finite everywhere including the asymptotic region $r \rightarrow \infty$.   
In the limit $a\rightarrow 0$, this solution for $\Phi$ will reduce to
that for the extremal case, given as
\begin{equation}
\Phi(r) = \left (\frac{C_1}{r} + \frac{C_4}{2}\right )^2 -1
\end{equation}
Here $C_1>0$ and $C_4>2$ is a requirement for positivity of the brane separation. 
Figure 1 shows the cases with $C_1=1$, $a=2$, $C_4=3$ (blue curve)
and $C_4=-3$ (red curve). It may be noted from the figure 
that for $C_4=-3$ the radion
ends up having zeros and is therefore such a choice of $C_4$ is not
permitted. 

\begin{figure}[h]
\begin{minipage}{18pc}
\includegraphics[width=18pc]{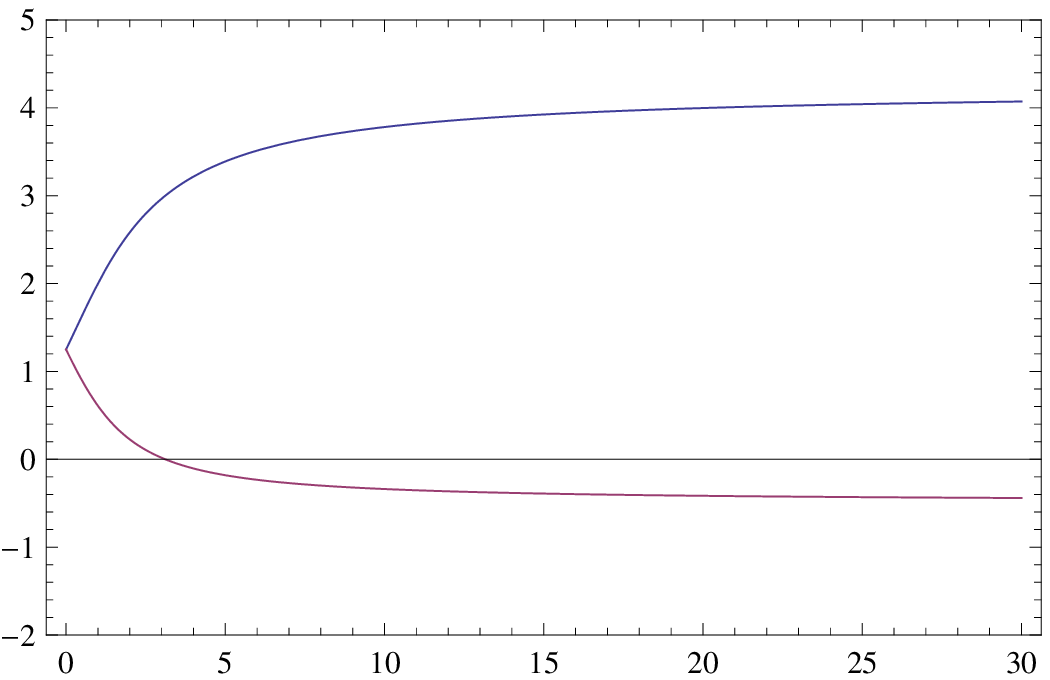}
\caption{$\Phi(r)$ vs. $r$ for $a^2>0$; $C_1=1, a=2, C_4=3 (blue), C_4=-3(red)$}
\end{minipage}\hspace{2pc}%
\begin{minipage}{18pc}
\includegraphics[width=18pc]{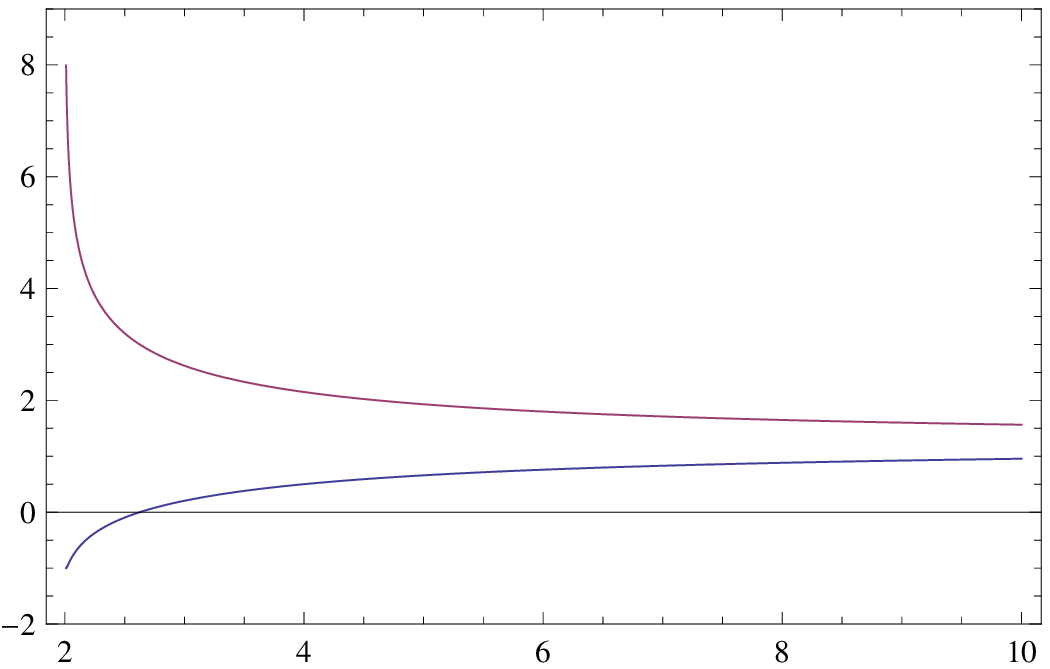}
\caption{$\Phi(r)$ vs. $r$ for $a^2<0$; $C_1=1, a=2, C_4=-3 (red), C_4=3 (blue)$}
\end{minipage}
\end{figure}

\noindent When $e^2< M^2$ (i.e. the black hole solution with horizon at
$r=a$, $a=\sqrt{-a^2}$), we obtain
\begin{equation}
\Phi (r) = \left (\frac{C_1}{2a} \ln \vert\frac{r-a}{r+a}\vert + \frac{C_4}{2}\right )^2
-1
\end{equation}
This solution is valid for all $r \geq a$. It may be observed that
at the location of the horizon, the inter-brane distance becomes
infinitely large (see Figure 2). As shown later, this divergence 
implies a divergent
matter stress energy at the horizon, thereby 
making the solution physically disallowed. Note also that for
$C_4=3$ the radion has zeros and therefore such a choice of the
parameters is not permissible. 
In the limit $a\rightarrow 0$ (from the $a^2<0$ side) limit, the
above solution does approach the one mentioned above for $a=0$. 

\begin{figure}[h]
\includegraphics[width=18pc]{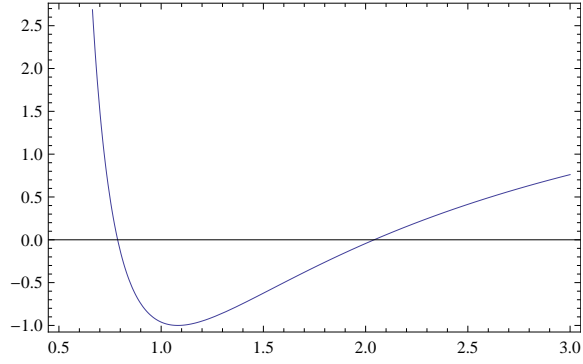}
\caption{$\Phi(r)$ vs. $r$ for Schwarzschild; $C_4=2, C_1=2, M=1$}
\end{figure}

\noindent Finally, let us obtain the radion for the Schwarzschild case, i.e.
when $e=0$. This turns out to be:
\begin{equation}
\Phi= \left (\frac{C_1}{M} \ln \frac{2r-M}{2r+M} + C_4\right )^2 -1
\end{equation}
One can easily show from the above expression that on the horizon at $r = \frac{M}{2}$, $\Phi$ diverges making the solution again
physically unacceptable.
We illustrate this behaviour of radion profile in Figure 3. We note that
apart from the divergence of the radion at the horizon, the radion profile
hits $\Phi=0$ at locations outside the horizon. This is a generic feature.
Coupled with the radion divergence at the horizon, one is forced to
conclude that a physically allowed radion is not possible in a
Schwarzschild spacetime.

\subsection{The Weak Energy Condition inequalities}

\noindent Using the expressions of $U$, $\Phi$ and $f$ mentioned in the previous subsection, 
in  Eqns. (9)-(11) we obtain non-zero $\rho$, $\tau$ and $p$, i.e.
we have non-trivial on-brane matter. 

\noindent It is now necessary to see if the matter is reasonable in terms
of satisfying the energy conditions \cite{wald}  and whether the 
$a\rightarrow 0$ limit yields our earlier result \cite{kls}. 

\noindent From the field equations we obtain,

\begin{eqnarray}
\frac{\kappa^2}{\ell} \rho = \Phi\frac{e^2}{U^2 r^4} + \frac{\alpha^2 M^2}{r^4 f^2}
-\frac{\Phi'}{U f} \left (-\frac{M}{r^2} + \frac{4a^2}{r^3}\left (1+\frac{M}{4r}
\right ) \right ) \\
\frac{\kappa^2}{\ell} \tau = -\Phi \frac{e^2}{U^2 r^4}
+\frac{3 \alpha^2 M^2}{r^4 f^2}  +\Phi' \left ( \frac{f}{Ur} + \frac{1}{r} - \frac{2 a^2}{r^3 f} \right )
\end{eqnarray}
where we have used $C_1=\alpha M$ where $\alpha$ is a proportionality constant.


\noindent The pressure $p$ can be obtained from the tracelessness condition. 
It is
given as,
\begin{eqnarray}
\frac{\kappa^2}{\ell} p = \frac{1}{2} \left ( \rho -\tau\right )
\end{eqnarray}
Recall that  $\rho\ge 0$, $\rho+\tau\ge 0$, $\rho+p\ge 0$
constitutes the Weak Energy Condition (WEC) and the subset $\rho+\tau\geq 0$,
$\rho+p\geq 0$ defines the Null Energy Condition (NEC)) \cite{wald}.
The analysis of the energy conditions will help us
know about the nature of the traceless matter that must be there
on the brane, if the solution has to exist.

\noindent It is useful to write down the expressions for $\rho+\tau$
and $\rho+p$ in a compact form. They are given as:
\begin{eqnarray}
\frac{\kappa^2}{\ell} \left (\rho+\tau\right )
= \frac{2}{f^2 r} \left [ \Phi' \left (1- \frac{a^4}{r^4} \right )
+\frac{2 \alpha^2 M^2}{r^3} \right ] \\
\frac{\kappa^2}{\ell} \left (\rho+p\right )
=\frac{1}{r^4} \left [ \frac{2\Phi e^2}{U^2} +\frac{\Phi'}{2U f}\left \{r^3(U^2-3f^2)- r a^2\right \}\right ] 
\end{eqnarray}

\noindent We now analyse the various cases separately.

\noindent {\bf Case 1 ($\alpha=1,e^2>M^2$):} 
In Figs. 4,5,6 we have chosen $e=5,M=3$ (naked singularity 
ar $r=1$) so that
$e^2-M^2=16$. We also choose $\alpha=1$ which means that the $M$ in the
metric functions is the same as the $C_1$ in the radion field solution. 
The plot of $\rho$ vs. $r$ (Figure 4) demonstrates that $\rho$
is indeed positive over the entire domain of $r$. Also, from the 
expression for $\rho$, it is clear that the dominant term
which varies as $\frac{1}{r^2}$ as $r\rightarrow \infty$ has a positive
coefficient. 
However, the  $\rho+\tau\geq 0$ inequality is violated in a
finite region around $r=1$ (Figure 5). In the same way, we note that
the $\rho+p\geq0$ inequality is violated for large $r$, a fact
which we demonstrate in Figure 6. In Fig. 6, the $y$-axis is
scaled by a factor of ${10}^{7}$ and we plot from $r=100$ to
$r=1000$. The negativity of $\rho+p$  at large values of $r$, 
is evident from this figure.
Thus the on-brane matter must necessarily violate the WEC
if the naked singularity is a viable solution.
\begin{figure}[h]
\begin{minipage}{12pc}
\includegraphics[width=12pc]{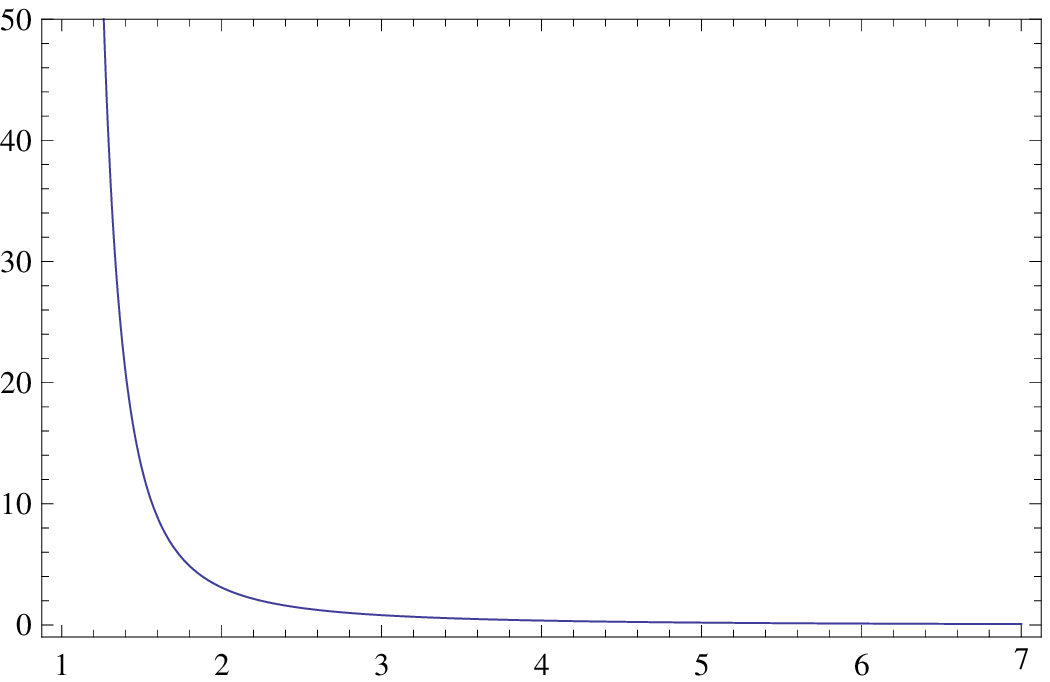}
\caption{$M=3, e=5, C_4=3$, $\alpha=1$, $\rho$ vs. $r$.}
\end{minipage}\hspace{2pc}%
\begin{minipage}{12pc}
\includegraphics[width=12pc]{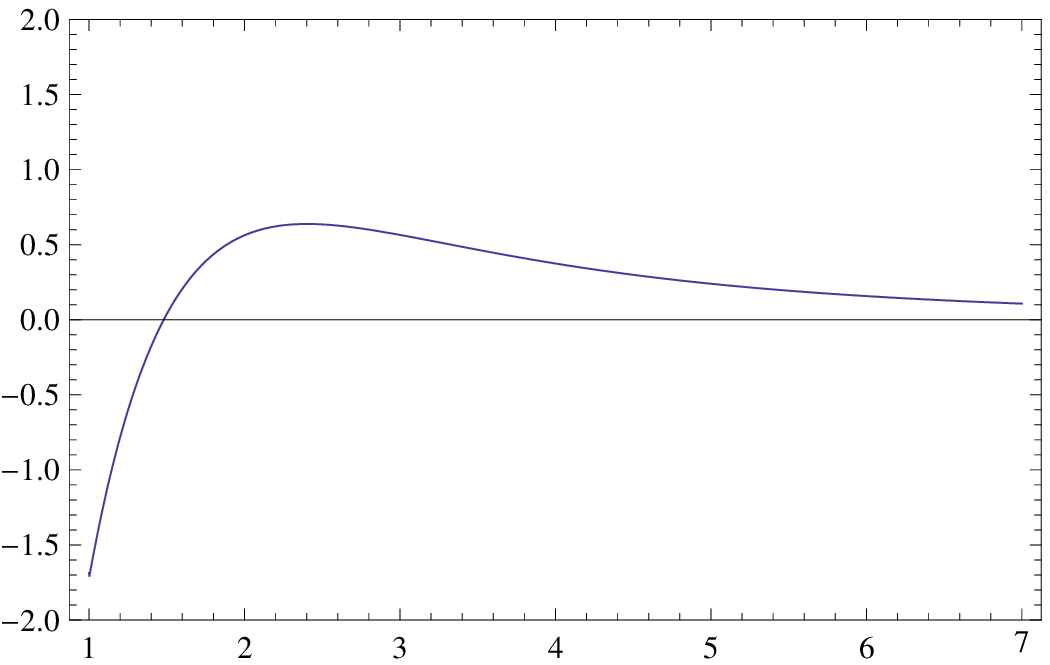}
\caption{$M=3, e=5, C_4=3$, $\alpha=1$, $(\rho+\tau) $ vs. $r$.}
\end{minipage}
\begin{minipage}{12pc}
\includegraphics[width=12pc]{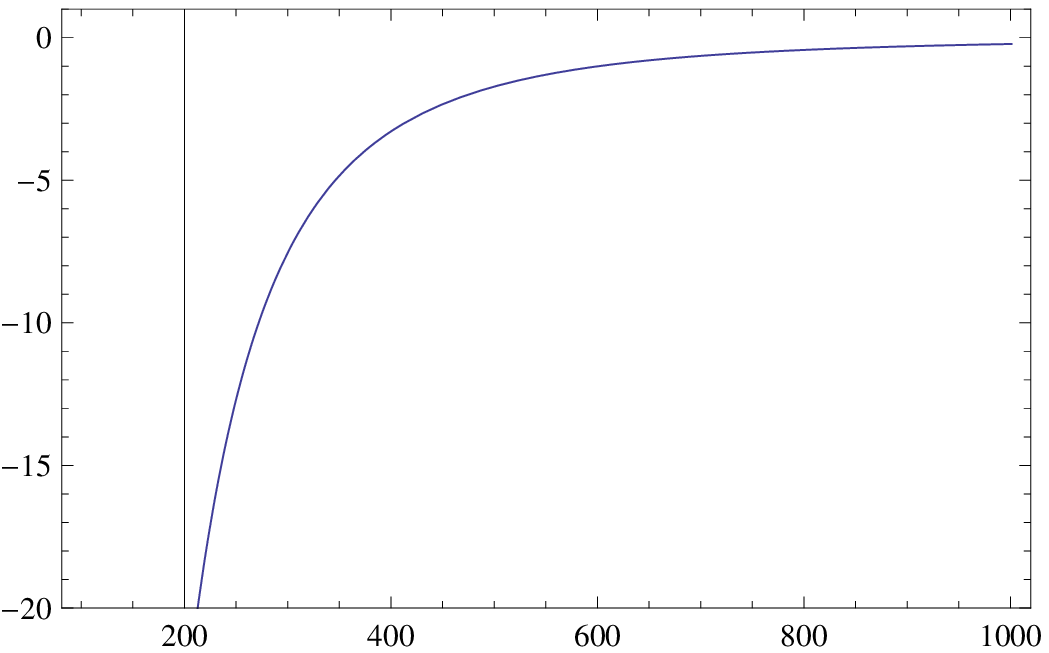}
\caption{$M=3, e=5, C_4=3$, $\alpha=1$, ${10}^{7}(\rho+p)$ vs. $r$.}
\end{minipage}
\end{figure}

\noindent To see this more explicitly let us go back to the
compact expression for $\rho+\tau$ quoted in (25), with $\alpha=1$. 
Here, note that the second term is positive but the positivity of the
first term depends crucially on the sign of $\Phi'$ as well as the 
value of $r$. 
Now, we can have a solution $\Phi$ (for $e^2-M^2>0$), with an always positive  
$\Phi'$ (see Figure 1, blue curve) which, in turn will ensure 
the positivity of the brane separation. Further, note that in this case, 
$f$ is never zero. 
Hence the $\rho+\tau\geq 0$ inequality is
never violated as long as $r>a_{crit}$. For $r<a_{crit}$
there is a violation. The value of $a_{crit}$ is determined by the
$r$ for which the term in square brackets in (25) turns negative.
$a_{crit}<a$, as seen in Figure 5. In general, the value of
$a_{crit}$ can be found from a solution of the
transcendental equation:
\begin{equation}
tan^{-1}\frac{a_{crit}}{a}+\frac{C_4 a}{2M} =\frac{a a_{crit}}{a^2-a_{crit}^2}
\end{equation}
Further, it is easy to see that near the naked singularity one cannot
avoid a violation of the WEC by any choice of the parameters. Let us
evaluate the term in square brackets in (25) at $r=\frac{e-M}{2}$ (the 
location of the naked singularity). One finds that
\begin{equation}
[....]= \frac{16}{M(\nu-1)^3} \left [ 1-\frac{C_4}{2}-\frac{2}{\sqrt{\nu^2-1}}
\tan^{-1}\sqrt{\frac{\nu-1}{\nu+1}}\right]
\end{equation}
where $\nu=\frac{e}{M}>1$. Note, that earlier we 
found $C_4>2$ from the requirement of positive brane separation.
Thus the R. H. S. of (28) is always negative for any $\nu>1$ and
$C_4>2$. One can control the amount of violation by increasing $M$
(since it appears in the denominator) such that $M^2 - e^2$ is negative. 
Similarly, by adjusting $C_4$, $M$, $e$ one can control the extent
of the region (the value of $a_{crit}$ where the WEC will be violated).
But there is no way to avoid the violation of the $\rho+\tau$ inequality
though it may be less in value or confined to a small region. 

\noindent In a similar manner, the $\rho+p$ inequality must necessarily
be violated for large $r$. The large $r$ limit of the $\rho+p$
expression clearly shows this feature. We have demonstrated this
violation of the $\rho+p$ WEC inequality in Figure 6.  

\noindent{\bf Case 2 ($e^2-M^2>0,\alpha\neq 1$):}  
In contrast, if we choose $\alpha\neq 1$ (i.e. $M\neq C_1$), 
the Eqn. (29) becomes
\begin{equation}
[....]= \frac{16\alpha}{M(\nu-1)^3} \left [ \alpha-\frac{C_4}{2}-\frac{2\alpha}{\sqrt{\nu^2-1}}
\tan^{-1}\sqrt{\frac{\nu-1}{\nu+1}}\right]
\end{equation}
Here, with appropriate choices of $\alpha$, $C_4$ and $\nu$ one can satisfy the
$\rho$, $\rho+\tau$  inequalities over the required domain, i.e. from
$r=\frac{e-M}{2}$ to infinity. For example, with $M=6$, $e=10$, $\alpha=6$
and $C_4=3$
we find that in the domain $2 \leq r \leq \infty$ there is no violation
of the $\rho$, $\rho+\tau$ inequalities
(the naked singularity is at $r=2$). This is shown in Figures  7,8.
However, the $\rho+p$ inequality still remains violated at large $r$,
a fact we show in Figure 9. 

\noindent By tuning $\alpha$,
$C_4$ and $\nu$ one can move around and reduce the extent of WEC violation
though it cannot be avoided completely for the $\rho+p$ inequality. 
Thus, for a naked singularity, WEC violation of on-brane matter 
is necessary and this conforms
with the Cosmic Censorship Hypothesis \cite{wald}.

\begin{figure}[h]
\begin{minipage}{12pc}
\includegraphics[width=12pc]{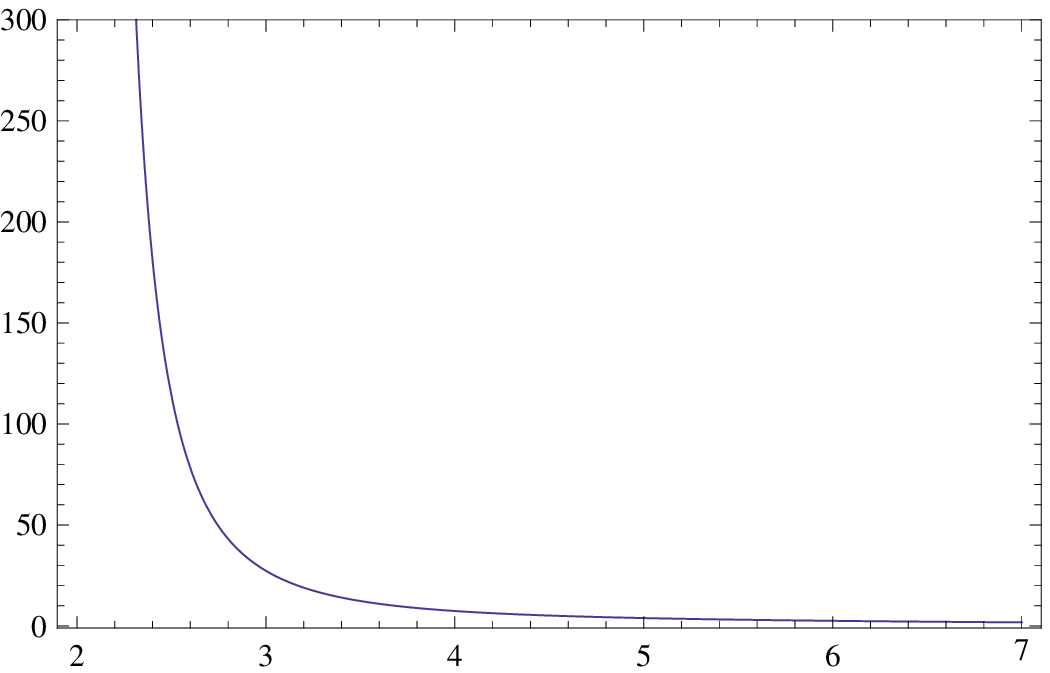}
\caption{$M=6, e=10, C_4=3$, $\alpha=6$, $\rho$ vs. $r$.}
\end{minipage}\hspace{2pc}%
\begin{minipage}{12pc}
\includegraphics[width=12pc]{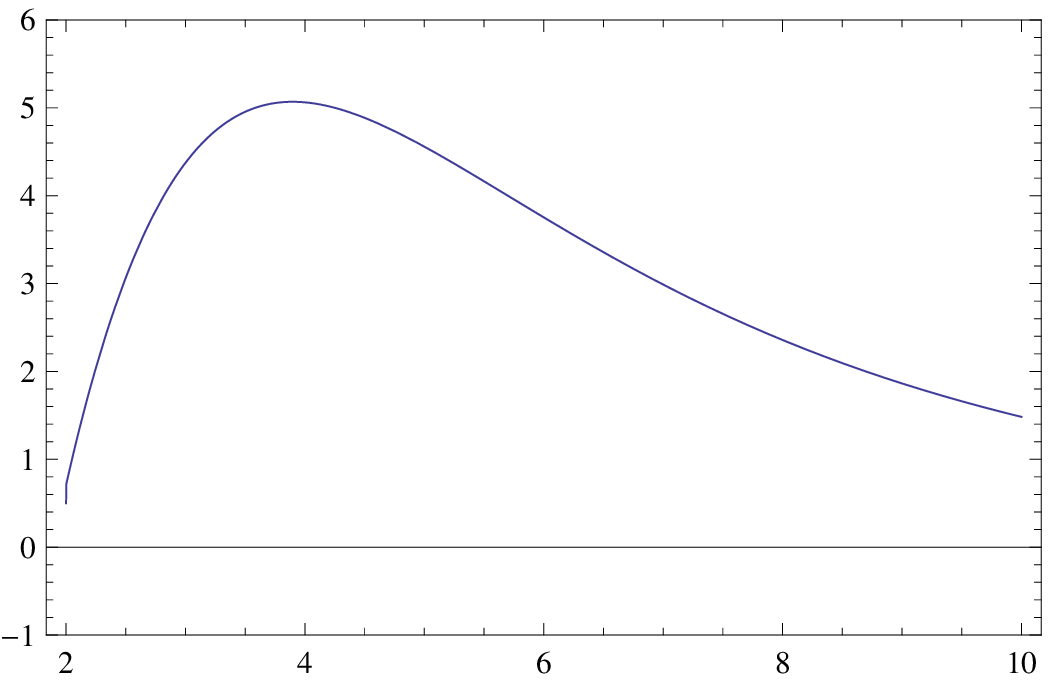}
\caption{$M=6, e=10, C_4=3$, $\alpha=6$, $(\rho+\tau) $ vs. $r$.}
\end{minipage}
\begin{minipage}{12pc}
\includegraphics[width=12pc]{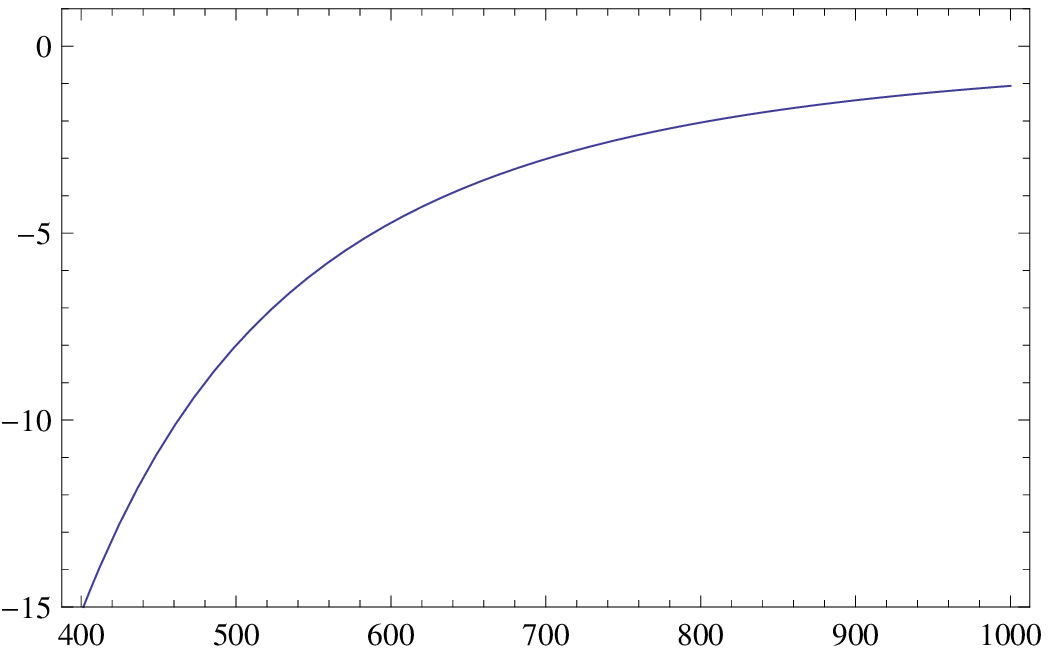}
\caption{$M=6, e=10, C_4=3$, $\alpha=6$, ${10}^{6}(\rho+p)$ vs. $r$.}
\end{minipage}
\end{figure}

\noindent Recall that when $e^2-M^2\leq 0$, the $\Phi$ diverges at
the black hole horizon $r=a$ 
which makes the RN black hole solution unphysical. 
The divergence of $\Phi$ also implies a divergence of $\rho$, $\tau$ and
$p$, as is evident from (22),(23). Thus, we do not discuss this case
any further here.

\noindent In our earlier paper, we had noted that the extremal
limit solution also does require on-brane matter which 
violates the energy condition inequality $\rho+\tau\ge 0$. 
Here we have seen that this violation persists for all $e^2> M^2$.

\section{Conclusion}

\noindent In this article, we have discussed the viability of the 
various well known GR solutions like Schwarzschild and Reissner--Nordstrom
in the context of the KS theory of gravity.
The crucial element in this work is
related to finding a stable radion which is finite and non-zero everywhere. 
The RN black hole solution requires an infinite inter-brane distance at
the horizon -- a fact which makes it unphysical. 
On the other hand, the Schwarzschild solution requires a radion which
diverges at the horizon, vanishes at two values of $r$ outside the horizon and
is negative between these values. The RN naked singularity and the extremal RN 
solution do have a
non-zero and finite radion, but our analysis shows that
the required on-brane matter 
violates the Weak and the Null Energy Condition. 
For the naked singularity, this feature conforms  with the Cosmic Censorship 
Conjecture \cite{wald}.

\noindent A possible way out of the problems mentioned here is to look for
solutions which are non-singular in nature and see if the radion is
finite and non-zero everywhere and the on-brane matter
satisfies the energy conditions. The finiteness of the radion seems to
be in conflict with the existence of a black hole horizon. Does this mean
that there are no eternal black hole solutions in this theory?   
We have not proved any such statement but the analysis on the
RN and Schwarzschild solutions seem to suggest such an outcome.

\noindent Finally, it may be useful to assume a specific form
of the on-brane matter (on either
or both branes),
and then find the radion and the metric functions. This will genuinely
be like {\em finding an exact solution}, given the matter content on the branes.
However, knowing the complicated 
nature of the field  equations, this will not be easy to do. 
Further, if we remove the traceless requirement, the equations will
become even more difficult to solve.

\noindent One might view the KS theory as a scalar-tensor theory in its
own right. Then, of course, the radion is just another scalar field
without any reference to braneworlds and it need not satisfy the
requirements we have mentioned in this paper. However, such an approach
is not the main motivation of this article where we have 
chosen to view the radion as related to the proper distance between branes 
located in a higher dimensional bulk spacetime.

\end{document}